# Plasmonic properties of individual gallium nanoparticles

Michal Horák[1,*], Vojtěch Čalkovský[1,2], Jindřich Mach[1,2], Vlastimil Křápek[1,2], Tomáš Šikola[1,2]

[1] Central European Institute of Technology, Brno University of Technology,

Purkyňova 123, 612 00 Brno, Czech Republic

[2] Institute of Physical Engineering, Brno University of Technology,

Technická 2, 616 69 Brno, Czech Republic

[*] corresponding author: michal.horak2@ceitec.vutbr.cz

______________________________________________________________________

**Abstract**

Gallium is a plasmonic material offering ultraviolet to near-infrared tunability, facile and scalable preparation, and good stability of nanoparticles. In our contribution, we experimentally demonstrate the link between the shape and size of individual gallium nanoparticles and their optical properties. To this end, we utilize scanning transmission electron microscopy combined with electron energy loss spectroscopy. Lens-shaped gallium nanoparticles with a diameter between 10 nm and 200 nm were grown directly on a silicon nitride membrane using an in-house developed effusion cell operated at ultra-high vacuum conditions. We have experimentally proved that they support localized surface plasmon resonances and their dipole mode can be tuned through their size from ultraviolet to near-infrared spectral region. The measurements are supported by numerical simulations using realistic particle shapes and sizes. Our results open the way for future applications of gallium nanoparticles such as hyperspectral absorption of sunlight in energy harvesting or plasmon-enhanced luminescence of ultraviolet emitters.

______________________________________________________________________

In metallic nanostructures, collective oscillations of free electrons are strongly coupled to the electromagnetic field forming the excitations called localized surface plasmons (LSP). A characteristic feature of LSP is a strong enhancement of electromagnetic field within the surrounding dielectric together with its confinement on the subwavelength scale, which can be utilized to control various optical processes even below the free space diffraction limit [1]. This feature is utilized in numerous applications [2]. The most common plasmonic metals are gold and silver, but their performance is restricted in lower wavelengths by interband transitions. Consequently, gold supports LSP at wavelengths longer than 550 nm and silver supports LSP above 350 nm. The ultraviolet and whole visible spectral range is covered by aluminum [3], magnesium [4], and gallium [5,6,7]. The ultraviolet plasmonic activity was theoretically studied and discussed also for bismuth, chromium, copper, indium, lead, palladium, platinum, rhodium, ruthenium, tellurium, tin, titanium, and tungsten [8,9,10]. Moreover, unconventional plasmonic materials are utilized in specific application fields, including the spectro-electrochemistry prospect of silver amalgam nanoparticles [11] or tunable plasmonic devices or metasurfaces made of phase-changing materials such as vanadium dioxide [12,13] or gallium [5,14,15].

Gallium is a metal with a melting temperature of 29.7 °C. It has several solid-state phases which enable a variety of phase-changing systems [5,14,16,17]. The volume plasmon energy of gallium is 13.7 eV [18]. Gallium does exhibit any pronounced interband transitions in a wide region from ultraviolet to infrared [19] which makes it an ideal plasmonic candidate. Further, gallium is non-toxic and rather friendly to the environment [20,21]. Gallium nanoparticles can be prepared by various bottom-up fabrication techniques like colloidal synthesis [22], optically regulated self-assembly [23], molecular beam epitaxy [24], and Joule-effect thermal evaporation [25]. Importantly, the low melting temperature of gallium allows low-temperature fabrication with low energy consumption. Prior studies reported plasmonic properties of gallium nanoparticles [6,26], tuning of plasmon resonance by oxidation creating Ga-$Ga_2O_3$ core-shell structure [27], gallium-indium alloy nanoparticles [28], and silver-gallium alloy nanoparticles [29]. There are numerous applications for such nanoparticles. They can be used, for example, as DNA biosensing platforms [30], for enhancing the luminescence of $MoS_2$ monolayers [31], for surface-enhanced Raman spectroscopy applications [32,33,34], or as anode material for Li-ion batteries [21].

Most of the studies of plasmonic properties of gallium addressed ensembles of nanoparticles. There are only two experimental studies focused on the optical properties of individual gallium nanoparticles. These studies utilized either electron energy loss spectroscopy [26] or cathodoluminescence [6]. Here, we present a study combining electron energy loss spectroscopy in a scanning transmission electron microscope (STEM-EELS) and numerical simulations to address the optical response of individual gallium nanoparticles. We show the spectral tunability of the dipole LSP mode over the near-infrared to the ultraviolet spectral range and correlate it with the size of gallium nanoparticles.

We have prepared gallium nanoparticles by direct deposition of gallium atoms onto a 50 nm thick silicon nitride membrane using an in-house-developed gallium effusion cell in ultra-high vacuum conditions (see Methods). For the EELS analysis, we selected two samples fabricated at different temperatures. Sample A grown at a higher temperature (320°C) contains nanoparticles with diameters in the range from 10 nm to 200 nm, whereas sample B grown at a lower temperature (290°C) contains nanoparticles with diameters in the range from 10 nm to 60 nm. The micrographs of both samples are shown in Figure 1. The nanoparticles are stable under electron beam illumination during the measurement. The three-dimensional morphology of the nanoparticles (Figure 1e-g) was determined by STEM-EELS. All the particles have a similar shape to a lens with an aspect ratio (defined as diameter-to-height ratio) mostly between 2.5 and 2.8 (Table 1). Consequently, our lens-shaped gallium nanoparticles are all geometrically similar.

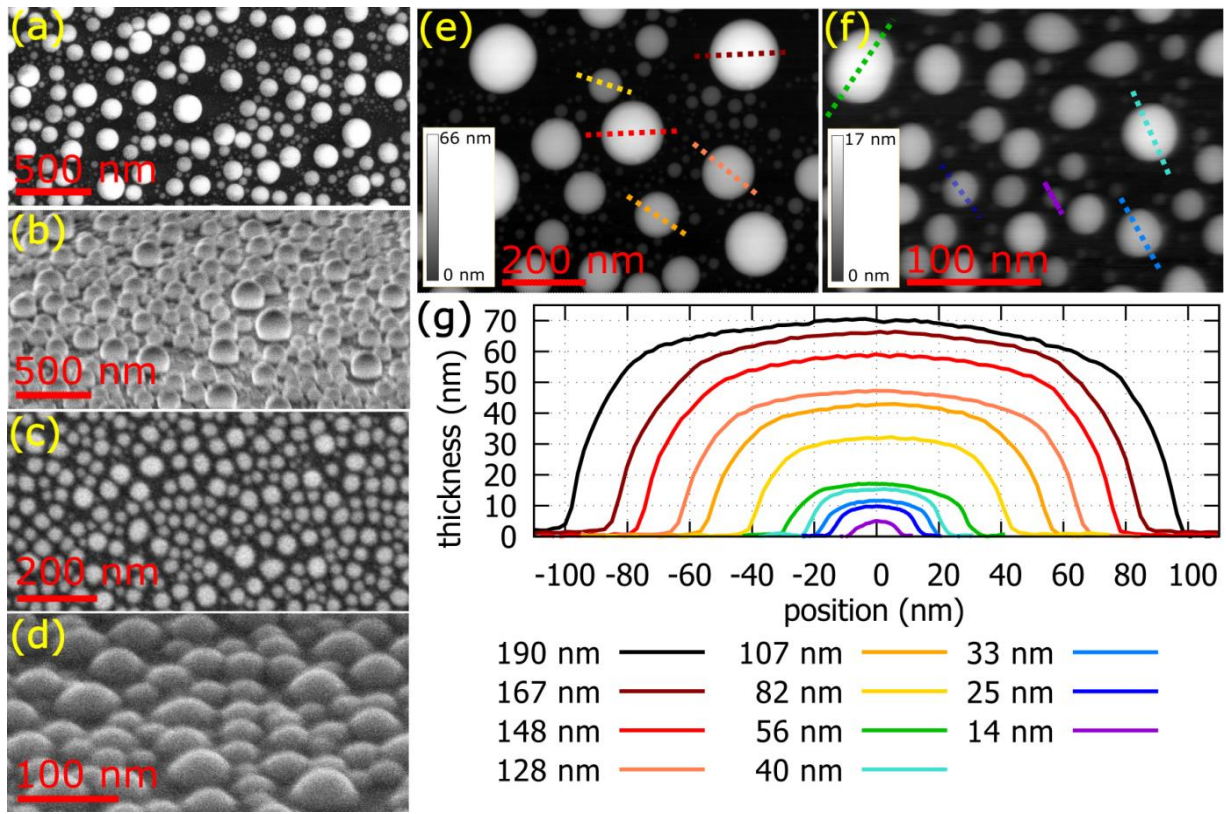

*Figure 1: Morphology of gallium nanoparticles: (a-d) SEM images of sample A with larger nanoparticles (a,b) and sample B with smaller nanoparticles (c,d) at zero (a,c) and 60° (b,d) tilt of the sample; (e,f) thickness maps measured by STEM-EELS for the sample A (e) and B (f); (g) thickness profiles of gallium nanoparticles in (e,f) with the diameters ranging from 14 nm (violet) to 190 nm (black). Note that the largest gallium nanoparticle is not shown in panel (e).*

First, we studied the crystallography and chemical composition of our gallium nanoparticles (Figure 2). According to the gallium phase diagram for nanoparticles or thin films [14,35], one may expect that such nanoparticles should be crystalline, consisting of the gamma phase of gallium (γ-Ga). However, the selected area diffraction pattern (Figure 2b,c) shows that our nanoparticles are amorphous or (partially) liquid. This observation can be attributed to a different fabrication process which is in our case done at the temperature of 290°C or 300°C whereas the gallium phase diagrams in the literature are for low-temperature depositions done at temperatures below 30°C [14,35]. The liquid phase of the nanoparticles is reported in Refs. [36,37] at a temperature above 254 K or 256 K, respectively.

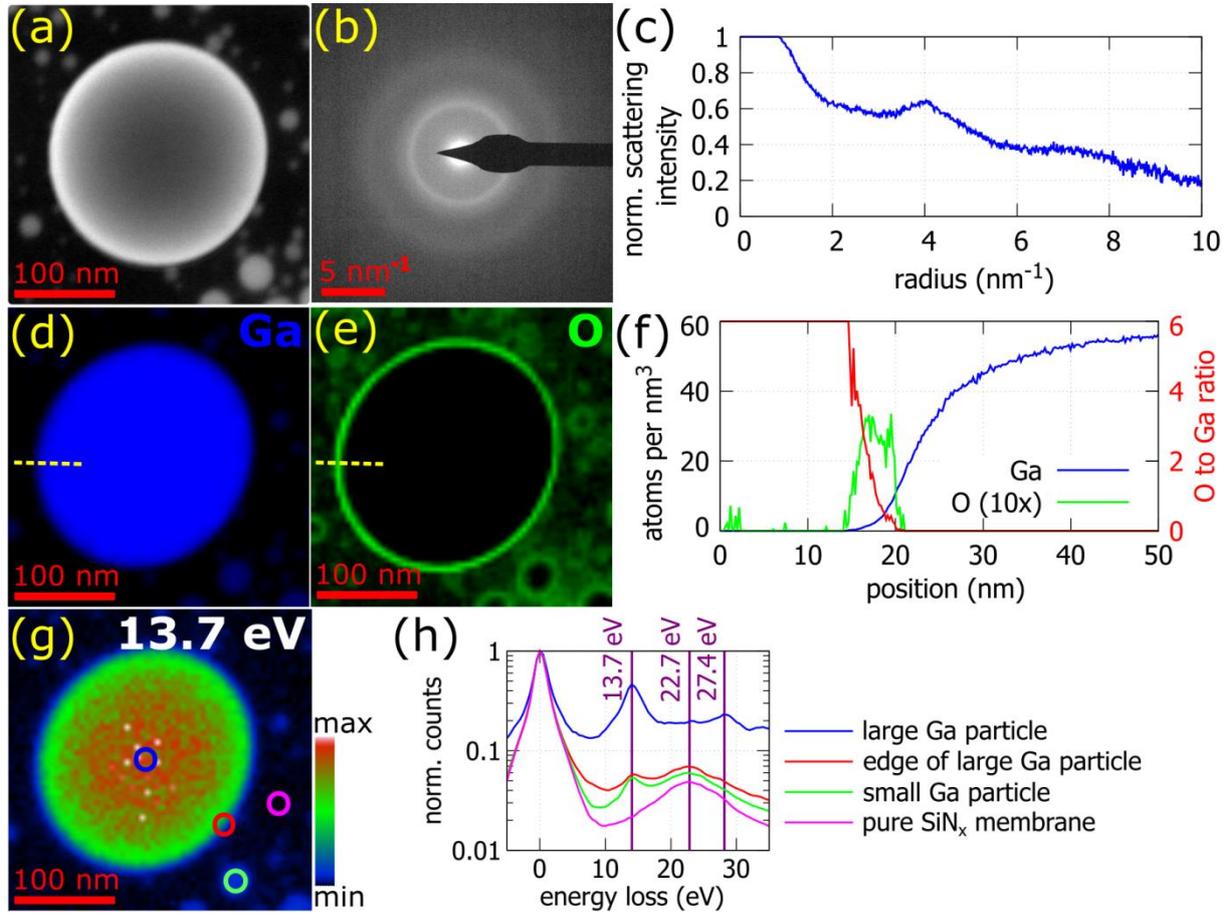

*Figure 2: Crystallography and chemical composition of a gallium nanoparticle: (a) STEM high-angle annular dark field (HAADF) micrograph of the nanoparticle; (b,c) selected area diffraction pattern (b) and its rotational average (c) of this nanoparticle proving its amorphous character; (d,e) chemical composition of this nanoparticle determined by STEM-EELS for the two most important elements Ga (d) and O (e); (f) line scans of the volumetric density of Ga and O (multiplied by 10) and their ratio over the dashed yellow lines in panels (d,e); (g) EELS map at the energy of (13.7 ± 0.5) eV corresponding to the volume plasmon of Ga; (h) low-loss EEL spectra at four different positions marked in (g).*

Chemical composition was determined by both core-loss and low-loss STEM-EELS (Figure 2d-h). The nanoparticles contain pure gallium in their volume. Oxygen is another significant element present in the vicinity of the nanoparticles, in particular, next to their edges. We will discuss two possible scenarios for the role of the oxygen: (i) a thin self-terminated gallium oxide layer encapsulating the nanoparticles, reported also in Ref. [6], (ii) contamination of the membrane, with the deposited gallium acting as a surface oxygen cleaner forming the oxygen-rich areas just in the very close vicinity of Ga nanoparticles. As oxygen is mostly concentrated next to the edges of the nanoparticles and partially on the membrane between the nanoparticles, it acts rather as a part of contamination of the membrane's surface than as a shell part of a core-shell Ga-GaO$_x$ structure. If a GaO$_x$ layer would be formed, it will be noticeable with a constant oxygen-to-gallium ratio, equal to 1.5 for the most usual stoichiometry Ga$_2$O$_3$, over several nm at the edge of the nanoparticle which is not the case of Figure 2f. Moreover, a GaO$_x$ layer would be detected by a shift of volume plasmon peak at 13.7 eV for Ga to a different energy for GaO$_x$ which is not the case in our low-loss STEM-EELS analysis (Figure 2g-h). Instead, we see a perfect match between the Ga elemental map (Figure 2d) determined by core-loss EELS and the energy-filtered map at 13.7 eV (Figure 2g), the energy of the Ga volume plasmon peak. In Figure 2h we present the low-loss STEM-EELS spectra from 4 different positions on the sample: the middle of the large gallium particle, the edge of this particle, a small gallium particle, and a pure silicon nitride membrane. The first three spectra contain a peak at 13.7 eV corresponding to the Ga volume plasmon peak. All four spectra contain a peak at 22.7 eV which corresponds to the SiN$_x$ volume plasmon peak coming from the membrane. And finally, the spectrum recorded in the middle of the large gallium particle contains a peak at 27.4 eV corresponding to the second volume plasmon peak of Ga (note that 2 · 13.7 eV = 27.4 eV). As there is no other peak visible in the low-loss EEL spectra, no other compound, except contamination, is present in the sample. Consequently, the second scenario is correct for our case and we have our gallium nanoparticles are homogeneous over the whole volume and composed of gallium of high purity.

Second, we have focused on the plasmonic properties of gallium nanoparticles which are measured by STEM-EELS at room temperature. The full low-loss EEL spectrum contains the contribution of multiple scattering mechanisms and also a large amount of the electrons transmitted through the sample without scattering (so-called zero-loss peak). To isolate the contribution of the LSP and remove all the other effects from EEL spectra, we define the signal spectrum as the full spectrum from which a reference spectrum is subtracted. The full spectrum is taken from the nanoparticle and the reference spectrum is from a pure membrane. The example of the full, reference, and signal EEL spectra is shown in Figure 3. We take the raw spectrum (red) integrated over the red area in the inset and subtract the spectrum of a pure membrane (blue) to get the signal (green) which consists of three peaks in the low-loss region. The peak at 13.7 eV corresponds to the volume plasmon peak of gallium [18]. The other two peaks are fitted by a Gaussian. At 7.4 eV we observe the peak corresponding to the surface plasmon (SP) of gallium. At energies below this value, we find peaks corresponding to individual localized surface plasmon (LSP) modes. In the spectra in Figure 3, we see the peak at 3.32 eV which corresponds to a dipole mode of LSP. This is the only peak that is supposed to change its spectral position when changing the nanoparticle size. As the properties, i.e., the energy and the full width at half-maximum (FWHM), of the volume plasmon peak and surface plasmon peak remain constant we can apply a fitting procedure, illustrated in Figure 3, which enables us to determine the energy of dipole LSP modes for the smallest nanoparticles even when they overlap with the surface plasmon peak. By fitting the spectral profile of the modes by a Gaussian, we obtained the LSP resonance energy E, and Q factor, defined as the LSP resonance energy divided by its FWHM to provide a complete analysis.

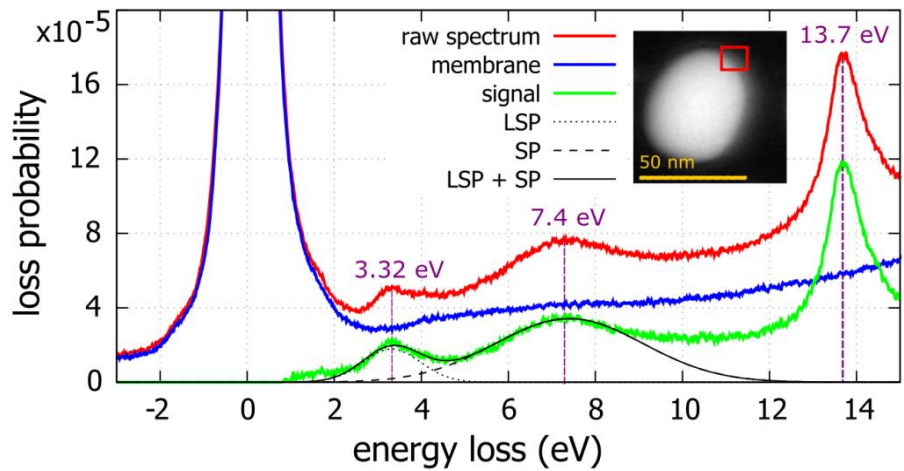

*Figure 3: Processing of the EEL spectrum in the case of gallium nanoparticle with a diameter of 56 nm. The low-loss EEL spectrum contains three peaks. The peak at 13.7 eV corresponds to the volume plasmon peak of gallium. At 7.4 eV we observe the peak corresponding to the surface plasmon (SP) of gallium. At energies below this value, we find peaks corresponding to individual localized surface plasmon (LSP) modes. In this spectra, we see the peak at 3.32 eV which corresponds to a dipole mode of LSP.*

Such a complete experimental analysis of the representative nanoparticles, whose morphology is discussed in Figure 1, is shown in Figure 4. According to the expectations, the EEL spectra should contain one peak at constant energy of 7.4 eV corresponding to the surface plasmon of gallium and one, or for larger particles more, peaks which should redshift with increasing nanoparticle size corresponding to the dipole and for larger particles to higher-order modes. This is exactly what Figure 4c shows. The dashed lines are a guide for the eyes. The first one follows the dipole LSP mode whose energy changes from 1.73 eV for the largest (190 nm in diameter) nanoparticle to 5.32 eV for the smallest (14 nm in diameter) nanoparticle as a function of the particle diameter. The second one follows the quadrupole LSP mode, which is visible just for larger particles. The third one follows the surface plasmon in gallium at constant energy of 7.4 eV. Consequently, we have experimentally proved that the dipole LSP mode of gallium nanoparticle is tunable from the ultraviolet spectral region, represented by the nanoparticle with a diameter of 14 nm with the dipole resonance at 5.32 eV (corresponding wavelength 233 nm), to the red end of the visible spectral region, represented by the nanoparticle with a diameter of 190 nm with the dipole resonance at 1.73 eV (corresponding wavelength 717 nm). Larger gallium nanoparticles would probably support LSP resonances in the infrared spectral region. We note that our results are in agreement with a similar system studied with different methods (finite-difference time-domain numerical simulations with a partial experimental verification by cathodoluminescence) described in the literature [6] where the dipole mode in gallium nanoparticles on a silicon substrate was found to be in the wavelength range from 200 nm to 800 nm depending on the particle size. The energy and the Q factor of the dipole LSP mode for all studied nanoparticles are summarized in Table 1. The Q factors are for most nanoparticles around 2. Naturally, these values are influenced by instrumental broadening, but still, we can compare them with Q factors found in the literature. Q factors obtained by the same analytical method (i.e., STEM-EELS) were evaluated for gold nanorods reaching values around 3 for the dipole mode at the resonant energy around 1.1 eV [38].

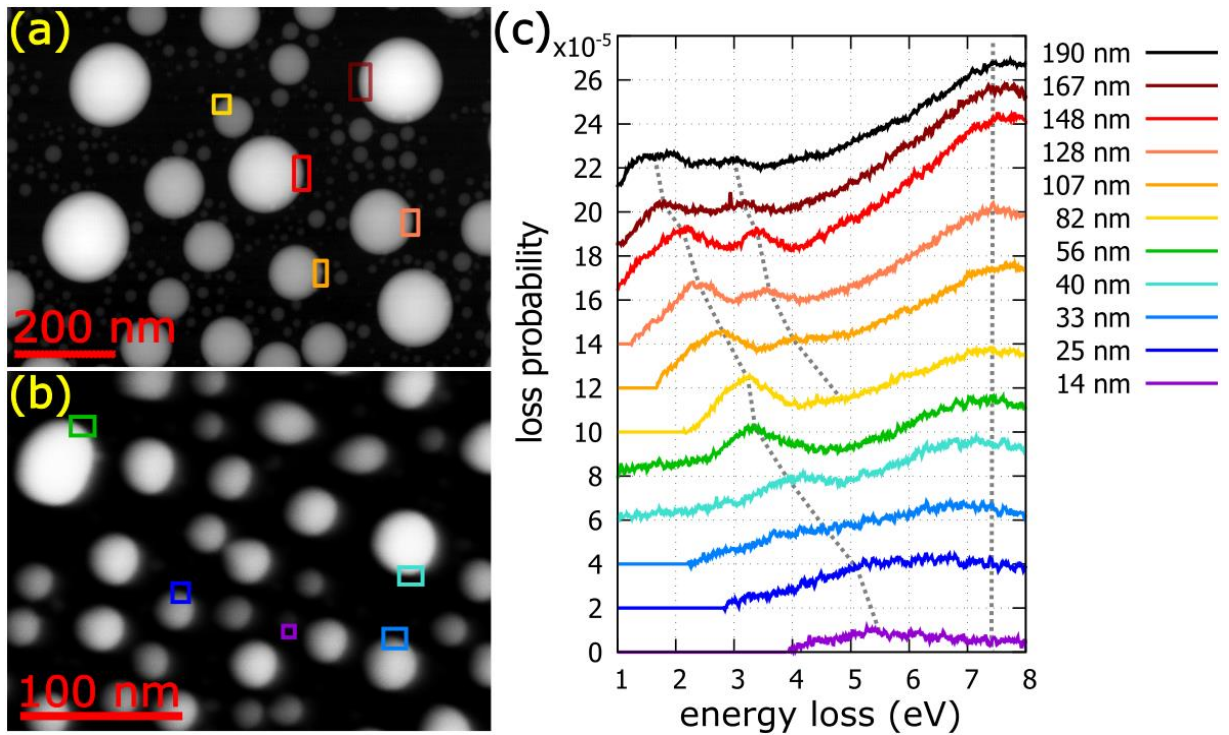

*Figure 4: Processed EEL spectra of gallium nanoparticles: (a,b) STEM-HAADF micrograph of the sample A (a) and B (b); (c) membrane subtracted EEL spectra integrated over marked areas in (a,b). Note that the largest gallium nanoparticle is not shown in panel (a). Dashed lines in (c) are guides for eyes and follow the dipole and the quadrupole localized surface plasmon mode whose energy changes as a function of the particle diameter and the surface plasmon in Ga at constant energy of 7.4 eV. The morphology of these gallium nanoparticles is shown in Figure 1.*

*Table 1: Structural and plasmonic properties of gallium nanoparticles.*

| | Diameter [nm] | Maximum thickness [nm] | Aspect ratio | Dipole LSP mode | | |
|---|---|---|---|---|---|---|
| | | | | Energy [eV] | Wavelength [nm] | Q-factor |
| Sample A | 190 | 71 | 2.7 | 1.73 | 717 | 1.4 |
| Sample A | 167 | 66 | 2.5 | 1.82 | 681 | 1.7 |
| Sample A | 148 | 59 | 2.5 | 2.13 | 582 | 1.5 |
| Sample A | 128 | 47 | 2.7 | 2.43 | 510 | 1.9 |
| Sample A | 107 | 42 | 2.5 | 2.78 | 446 | 1.9 |
| Sample A | 82 | 32 | 2.6 | 3.20 | 387 | 3.1 |
| Sample B | 56 | 17 | 3.3 | 3.32 | 370 | 2.1 |
| Sample B | 40 | 15 | 2.7 | 3.98 | 311 | 2.1 |
| Sample B | 33 | 12 | 2.8 | 4.44 | 279 | 1.5 |
| Sample B | 25 | 10 | 2.5 | 5.08 | 244 | 2.0 |
| Sample B | 14 | 5 | 2.8 | 5.32 | 233 | 3.1 |

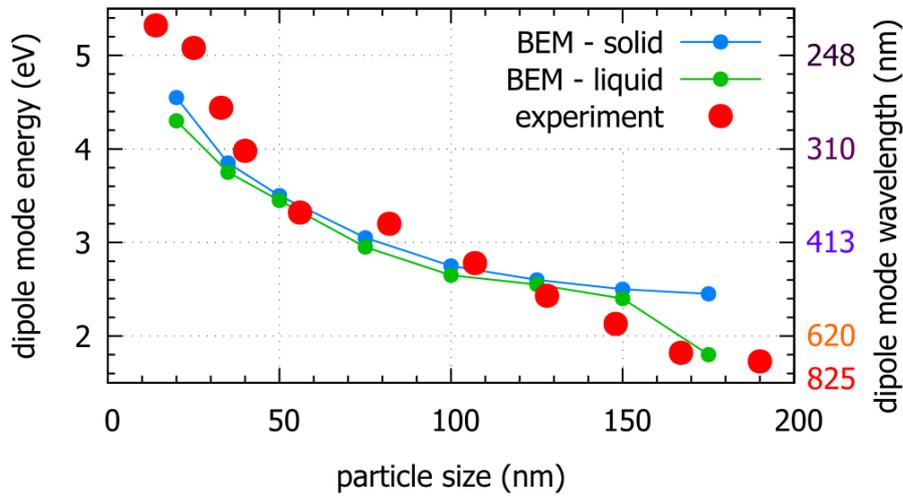

*Figure 5: Energy of the dipole localized surface plasmon mode in a set of gallium nanoparticles covering the spectral region from ultraviolet across visible to near-infrared.*

Additionally, we have performed numerical simulations to support our experimental results using the real particle sizes as the model parameters. We have performed two sets of calculations: one for liquid gallium and one for solid gallium using the dielectric functions taken from Knight et al. [6]. We note that the largest differences in the dielectric functions of different Ga phases are below 2 eV, where some of the Ga phases have interband transitions [17]. Figure 5 shows the energy of the dipole LSP mode in a set of gallium nanoparticles as a function of the nanoparticles' diameter. We see generally a good agreement between the experiment and the numerical simulations. However, if we consider the largest nanoparticles, we see a nonnegligible difference between the simulations for solid and liquid gallium. As our experimental values are closer to the curve for the liquid gallium, we expect that our nanoparticles are at least partially in the liquid phase. This finding is in line with the selected area diffraction pattern shown in Figure 2b and the corresponding discussion.

To conclude, we have explored the plasmonic properties of gallium nanoparticles at room temperature using scanning transmission electron microscopy combined with electron energy loss spectroscopy on a single particle level. Gallium nanoparticles with the size in the range between 10 and 200 nm were grown directly on the silicon nitride membrane by deposition of gallium atoms using an in-house-developed gallium effusion cell in UHV conditions. We have shown that their dipole mode can be tuned via their size from ultraviolet to visible spectral region. Our results are supported by numerical simulations using the real particle sizes as the model parameters. Concerning the potential applications, previous works reported that gallium nanoparticles can be used for enhancing luminescence or be used as a biosensing platform. Our results open the way for future applications such as hyperspectral absorption of sunlight in energy harvesting a plasmon-enhanced luminescence of ultraviolet emitters. Moreover, further understanding of their solid-to-liquid phase change will open a way for liquid plasmonics which would not suffer from quenching on grain boundaries and impurities and consequently might bring plasmonic nanoparticles with higher Q factors.

## Methods

**Gallium nanoparticle growth:** Samples with gallium nanoparticles were prepared by direct deposition of gallium atoms onto a 50 nm thick silicon nitride membrane using an in-house-developed gallium effusion cell in ultra-high vacuum (UHV) conditions. The size of the resulting nanoparticles is influenced by the temperature of the substrate, gallium flux, and time of the deposition. The samples were mounted on a bulk pyrolytic boron nitride (PBN) heater which allows us to carefully bring as constant temperature onto the whole membrane's surface as possible to control the size distribution of gallium droplets on the surface and their surface diffusion mobility. Sample A (larger nanoparticles) was deposited at 320°C for 3 hours. Sample B (smaller nanoparticles) was deposited at 290°C for 2 hours. The gallium atoms flux density was in both cases equal to $7.2 \cdot 10^{12}$ atoms·s$^{-1}$·cm$^{-2}$. Due to the fragility of the membranes, samples were left after deposition in UHV conditions for 2 hours to cool down and release the tension. Both samples were prepared at the pressure of $3.8 \cdot 10^{-8}$ Pa.

**Electron energy loss spectroscopy (EELS):** EELS measurements were performed with TEM FEI Titan equipped with a GIF Quantum spectrometer operated at the primary beam energy of 300 kV.

The chemical composition of the sample was determined by a STEM-EELS measurement in dual-EELS mode to acquire the low-loss and core-loss spectra simultaneously. The spatial resolution of the EELS spectrum images is determined by the pixel size, which was set to 5 nm for the EELS maps and 0.2 nm for the EELS line scan. Such settings led to acquiring one spectrum image with a stable electron beam in a reasonable time.

LSP resonances were measured in a monochromatic scanning regime. The beam current was set to 0.1 nA and the FWHM of the ZLP was around 0.15 eV. We set the convergence semi-angle to 10 mrad, the collection semi-angle to 14.4 mrad, and the dispersion of the spectrometer to 0.01 eV/pixel. These parameters were selected to acquire sufficient EELS signal over the areas with a large change in the thickness [39]. The acquisition time was adjusted to use the maximal intensity range of the CCD camera in the spectrometer and avoid its overexposure. The spatial resolution of the EELS spectrum images is determined by the pixel size, which was set to 2 nm. To reduce noise in the LSP signal, the EEL spectra were integrated over the rectangular areas at the nanoparticle edges (marked in Figure 4) where the LSP resonance is significant. They were further divided by the integral intensity of the ZLP to transform measured counts to a quantity proportional to the loss probability, membrane subtracted, and fitted by Gaussians (see Figure 3).

The thickness of the Ga nanoparticles was evaluated from the low-loss EELS in terms of relative thickness which is proportional to the absolute thickness with the inelastic mean free path (IMFP) as the constant of proportionality. The IMFP in gallium for the actual parameters of the electron beam (electron energy of 300 keV and collection semi-angle of 14.4 mrad) was calculated using the software package EELSTools by D. Mitchell [40] applying the algorithm of K. Iakoubovskii et al. [41] and equals to 156 nm. By multiplying the relative thickness by this value, we get the approximate absolute thickness of Ga nanoparticles.

**Simulations:** Numerical simulations of EEL spectra were performed using the MNPBEM toolbox [42] based on the boundary element method (BEM). The structures were modeled as gallium nanodiscs with a rounded upper edge and an aspect ratio of 2.5. The dielectric function of solid and liquid gallium was taken from Knight et al. [6] and the dielectric function of the silicon nitride membrane was set to 4 which is a common approximation in the considered spectral region [43]. The 300 keV electron beam was situated outside the nanodisc 5 nm far from its edge. In addition to the nanodisc model shape, we also performed simulations for the realistic shape of the structures. To this end, we utilized the thickness distributions shown in Figure 1, which were subsequently rotated to form particles with a cylindric symmetry. The energies of the LSP resonances corresponded rather well to those obtained for the model nanodisc shape, verifying thus its plausibility. At the same time, the realistic-shape simulations suffered from numerical instabilities, which led us to prefer the model shape.

**Acknowledgments**

This research was supported by Czech Science Foundation (22-04859S), MEYS CR under the project CzechNanoLab (LM2018110, 2020-2022), and Brno University of Technology (FSI-S-20-6485).